\documentclass[review]{elsarticle}

\usepackage{lineno,hyperref}

\usepackage[margin=1in]{geometry}

\pdfoutput=1

\usepackage{graphicx}
\usepackage{bm}
\usepackage{dcolumn}
\usepackage{setspace}
\usepackage{amsmath}
\usepackage{fancyhdr}

\usepackage{caption}
\usepackage{subcaption}

\modulolinenumbers[5]

\journal{arxiv}

\begin{document}

\begin{frontmatter}

\title{Effect of shape on mechanical properties and deformation behavior of Cu nanowires : An atomistic simulations study}


\author[mymainaddress1]{P. Rohith\corref{mycorrespondingauthor}}
\cortext[mycorrespondingauthor]{Corresponding author}
\ead{prohith@campus.technion.ac.il}


\author[mymainaddress2]{G. Sainath}

\author[mymainaddress3,mymainaddress4]{V.S. Srinivasan}

\address[mymainaddress1]{Nanomechanics Simulations Laboratory, Faculty of Mechnaical Engineering, Technion-Israel Institute of Technology, Haifa-320003, Israel}

\address[mymainaddress2]{Materials Development and Technology Division, Indira Gandhi Centre for Atomic Research, Kalpakkam, Tamilnadu-603102, India}

\address[mymainaddress3]{Homi Bhabha National Institute (HBNI), Indira Gandhi Centre for Atomic Research, Kalpakkam, Tamilnadu-603102, India}

\address[mymainaddress4]{Scientific Information Resource Division, Resource Management \& Public Awareness Group, Indira Gandhi Centre for Atomic Research, Kalpakkam, Tamilnadu-603102, India}


\begin{abstract}

We study the effect of nanowire shape on mechanical properties and deformation behaviour of Cu nanowires using atomistic simulations. 
Simulations were carried out on [100] nanowires with different shapes such as triangular, square, pentagon, hexagon and circular. 
Based on size, two different scenarios were considered. In first, all shapes have different surface area to volume ratio, while 
in second, all have the same surface area to volume ratio. The results indicate that the Young’s modulus is insensitive to shape 
in both cases. However, yield strength is different for different shapes. In both cases, triangular nanowire exhibit the lowest 
yield strength, while circular nanowire is the strongest. Deformation in all the nanowires is dominated by partials slip and 
twinning. Due to twinning, different shapes expose different surfaces at the twinned region. All nanowires show ductile failure 
and square nanowire exhibits the highest failure strain, while it is lowest for triangular nanowire. \\
  
\end{abstract}


\end{frontmatter}

\section{Introduction}

In recent years, nanowires have been a major focus of scientific research due to their remarkable electrical, 
magnetic, optical and mechanical properties \cite{Lieber2003}. These distinct properties of nanowires arise 
mainly due to their finite size, high surface area to volume ratio and limited defect density \cite{Review-1,Review-2}. 
With increasing size, the surface area to volume ratio decreases and becomes negligible for bulk materials. 
As a result, the size effects are often seen mainly in nanoscale materials. In view of potential applications 
in nano electro mechanical systems (NEMS), huge research on nanowires has been carried in the past decade to 
understand the mechanical behaviour under the influence of various  parameters \cite{Review-1,Review-2}. Out 
of all the parameters, the size of the nanowires has been considered as the most important parameter that 
influences the mechanical behaviour of nanowires \cite{Review-1,Review-2,Uchic,Peng2012,Yaghoobi2016}. It has 
been reported that the properties such as Young's modulus, yield strength and strain to failure are strongly 
dependent on nanowire size \cite{Review-1,Review-2,Rohith-Philmag,Rohith-CMS17}. With increasing nanowire size, 
the values of Young's modulus and yield strength first decreases and then saturates towards a constant value 
\cite{Rohith-Philmag,Rohith-CMS17,Zhu2012}. Zhu et al. \cite{Zhu2012} had performed in-situ tensile testing of 
Ag nanowires with diameters between 34 and 130 nm. It can be clearly seen from their studies that Young's modulus 
of nanowires decreases with increasing diameter and 
finally attains saturation at a magnitude corresponding to the Young's modulus of bulk Ag. Similarly, it has been 
reported that the yield stress also exhibit a rapid decrease in small size nanowires followed by gradual decrease 
towards saturation at larger size \cite{Rohith-Philmag}. Further, the yield stress value at saturation has been 
found to be close to the theoretical/ideal strength of the nanowire \cite{Rohith-Philmag}. Along with nanowire size, 
the other factors such as crystallographic orientation \cite{Park2006,Weinberger2012Review,Rohith-ComCondMater}, 
mode of loading \cite{Rohith-ComCondMater}, strain rate \cite{Xie2015}, temperature \cite{Xie2015,Cao2008,Sai-JAP} 
and shape \cite{Cao2008,Leach2007,Zhang2009} also influences the mechanical behaviour and deformation mechanisms 
of single crystalline nanowires.

With regard to nanowire shape, Leach et al. \cite{Leach2007} had performed the MD simulations on silver nanowires 
with different cross-sectional geometries such as pentagon, rhombic, and truncated-rhombic. It has been shown that 
the pentagon shape Ag nanowires exhibit higher strength compared to rhombic and truncated-rhombic shapes. They had 
also reported that for all the shapes, the yield strength decreases with increasing cross-section width \cite{Leach2007}. 
Similarly, Cao and Ma \cite{Cao2008} have shown that the yield stress of circular shaped $<111>$ Cu nanowires is 
about 10 GPa, while for square shaped nanowire 6.5 GPa has been reported as the yield strength. In the literature, 
it has been found that apart from mechanical properties, the nanowire shape influences many different material 
properties such as Debye temperature \cite{Arora2017}, melting mechanisms \cite{Kateb2018} and absorption\cite{Kadam2019}. 
For instance, Arora et al. \cite{Arora2017} had studied the role of shape and size on the Debye temperature of Au, Ag, 
Cu and Co with three different shapes like spherical, tetrahedral and octahedral. They had reported that Debye temperature 
is maximum for tetragonal shape, and minimum for spherical shape. They had explained this variation on the basis of 
surface area to volume ratio, which varies with shape and size \cite{Arora2017}. The surface area to volume ratio is 
minimum for spherical shape, while it is maximum for tetragonal \cite{Arora2017}. Kateb et al. \cite{Kateb2018} had 
reported the differences in melting mechanisms of Pd cub-octahedron and icosahedron structures. They had reported 
that the large icosahedron structures melts through a combination of surface and diagonal melting. On the other hand, 
in cub-octahedron structures, surface melting occurred by nucleation of the liquid phase at (100) planes and growth 
of liquid phase occurs at the surface before inward growth \cite{Kateb2018}. Using simulation studies, Kadam et al. 
\cite{Kadam2019} had studied the optical properties of GaAs quantum dots with different shapes such as cuboid, cylinder, 
dome, cone, and pyramid. They had suggested that the both cuboid and cylinder shaped quantum dots show maximum absorption, 
whereas minimum absorption is observed for the dome-shaped quantum dots \cite{Kadam2019}. All these studies indicate 
that the shape greatly influences various properties of materials. 

As the nanowire shape/geometry influences the surface area to volume ratio, which in turn affects the material properties, 
the mechanical behaviour of nanowires with different cross-sectional shapes such as triangular, pentagonal and hexagonal 
also should be known apart from the well studied square and circular shapes. In this context, it is surprising to find that 
there is no systematic study in the literature on how the nanowire shape influences the mechanical properties and associated 
deformation mechanisms of nanowires. Further, the details of twinning related effects such as the orientation of the twinned 
region and twin growth have not been investigated in nanowire shapes other than square and circular. In view of this backdrop, 
the present investigation is aimed at understanding the role of nanowire cross-sectional shape on mechanical properties, 
deformation and failure behaviour of nanowires using atomistic simulations. The tensile loading has been applied on nanowires 
with different shapes such as triangular, square, pentagonal, hexagonal and circular cross-sections.

\section{MD Simulation details}

The tensile deformation of Cu nanowires with different cross-section shapes have been performed using molecular dynamics 
(MD) simulations through Large scale Atomic/Molecular Massively Parallel Simulator (LAMMPS) package \cite{Plimpton-1995}. 
In order to specify the atomic description of Cu atoms, an embedded atom method (EAM) potential for FCC Cu given by Mishin 
and co-workers \cite{Mishin-2001} has been employed in the present study. This potential has been chosen for being able to 
reproduce the generalized stacking fault and twinning fault energies for Cu \cite{Liang-PRB}, which are key variables for 
predicting the dislocation and plasticity related properties. 

\begin{figure}[h]
\centering
\includegraphics[width=12cm]{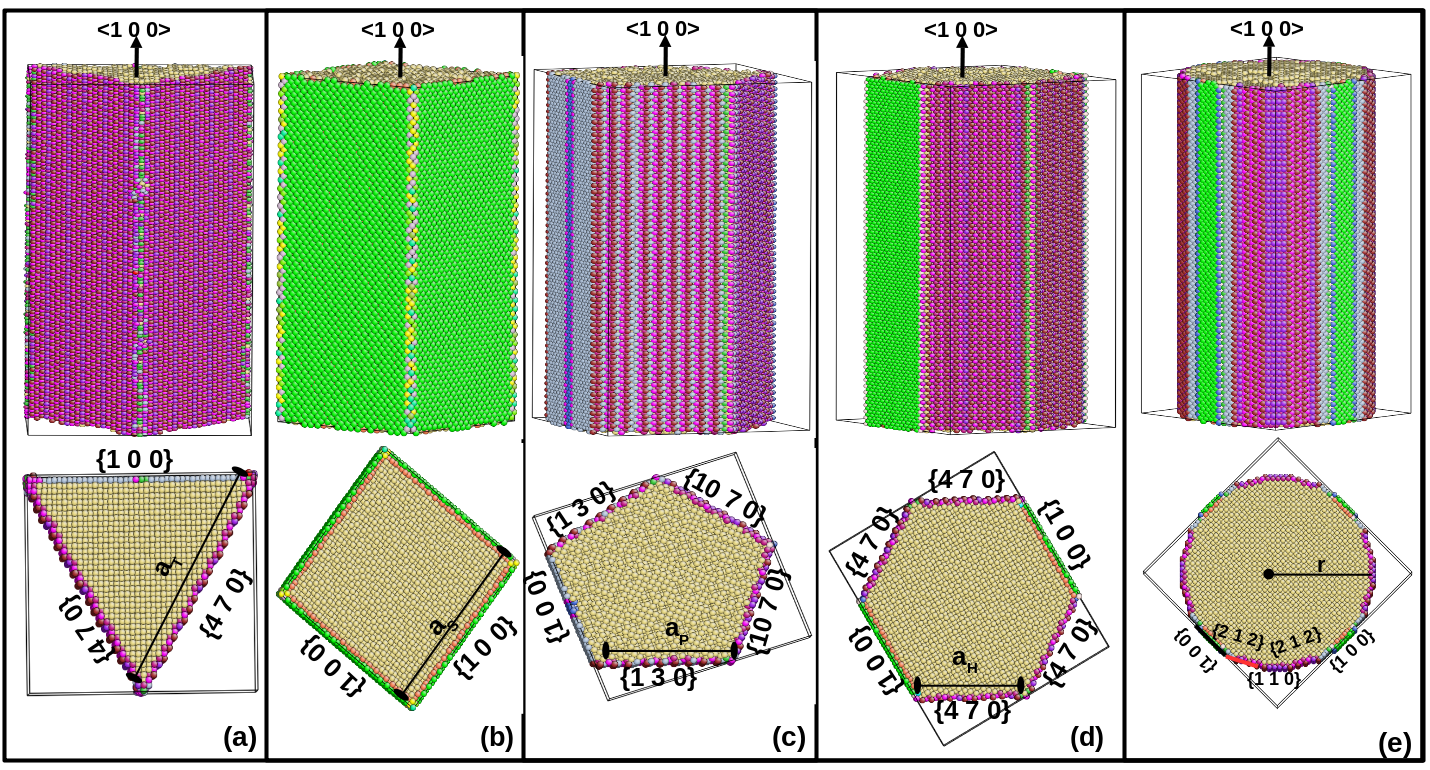}
\caption {The nanowires cross-section shapes considered in the present study. (a) Triangular, (b) square, (c) pentagonal, 
(d) hexagonal, and (e) circular, with their corresponding side surfaces.}
\label{initial-sidesurface}
\end{figure}  

In order to study the influence of shape of the nanowires, five different cross-sectional geometries i.e., triangle, square, 
pentagon, hexagon and circle have been chosen in the present study (Figure \ref{initial-sidesurface}). All the nanowires in the 
present study were oriented in $<$100$>$ along axial direction enclosed by different surfaces. The equilibrated side surfaces 
for all nanowires with different cross-sections has been shown in Figure \ref{initial-sidesurface}. It can be seen that the 
triangular and hexagonal shaped nanowires have enclosed by \{470\} and \{100\} type side surfaces, while square nanowire is 
enclosed by all \{100\} type surfaces (Figure \ref{initial-sidesurface}). Similarly, the pentagonal shaped nanowire has \{130\}, 
\{10 7 0\} and \{100\} as side surfaces, while the circular nanowire is surrounded by \{100\}, \{110\} and \{212\} type side 
surfaces (Figure \ref{initial-sidesurface}). For all the nanowires, periodic boundary conditions have been chosen only along the 
length direction ($<$100$>$), while the other directions were kept free in order to mimic an infinitely long nanowire. 

\begin{figure}[h]
\centering
\includegraphics[width=6cm]{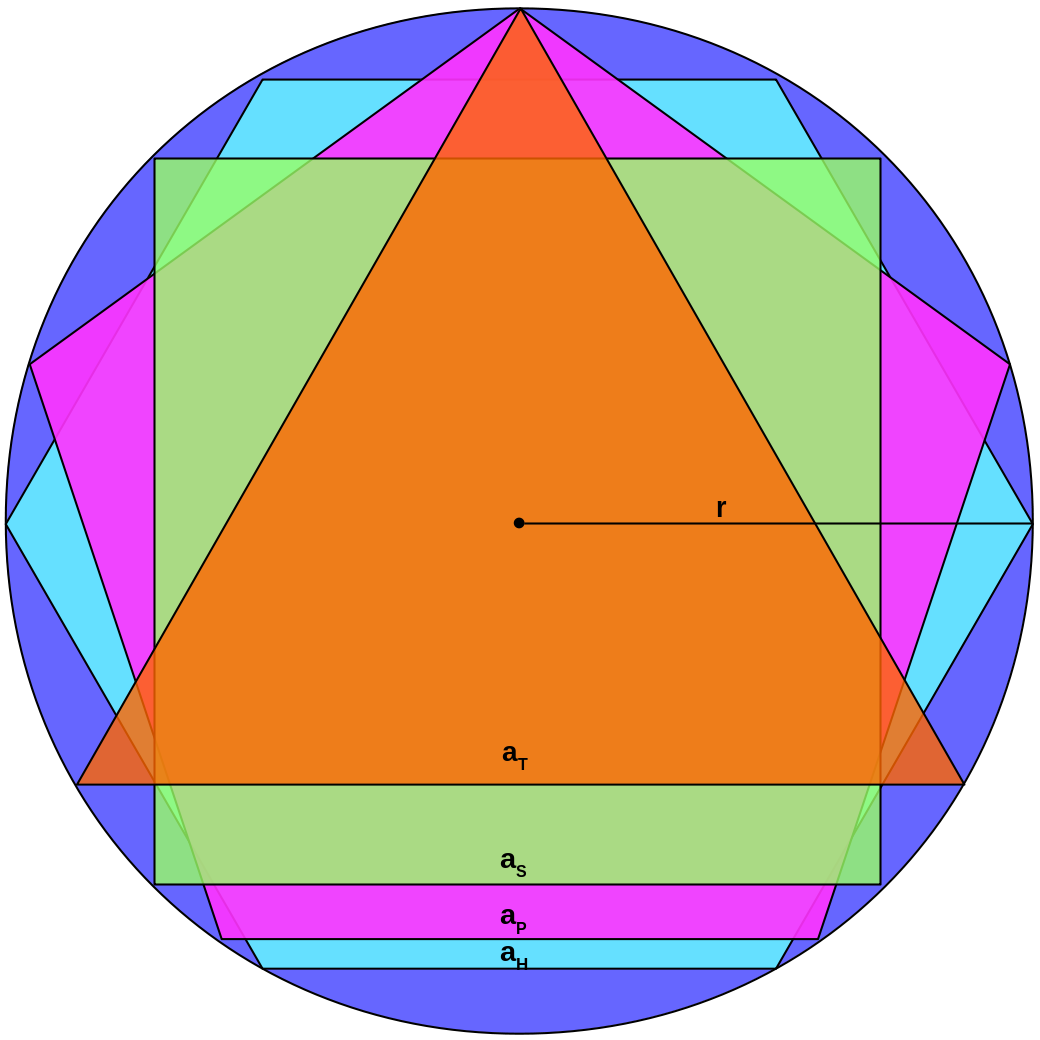}
\caption {A model system showing the inscription all the shapes within the circular shape. In the first scenario, the size for 
different shapes is chosen in such a way that all shapes can be inscribed in circular nanowire of radious 'r'. $a_T, a_S, a_P$, 
and $a_H$ represents the side length of triangular, square, pentagonal and hexagonal shaped nanowires.}
\label{Fig01}
\end{figure}

The size of nanowires of different shapes has been chosen in two different ways. In the first scenario, the size of the nanowires 
of different shapes is chosen in such a way that it can be inscribed in a nanowire with circular cross-section as shown in Figure 
\ref{Fig01}. As a consequence, all shapes have different surface area to volume ratio as shown in Table \ref{Dimensions}. In the 
second scenario, the size of the nanowires has been chosen in such a way that the surface area to volume ratio of all the shapes 
remains constant (Table \ref{Dimensions1}). The corresponding size of nanowires of different shapes is shown in Table \ref{Dimensions} 
and \ref{Dimensions1} for these two different cases of varying and constant surface area to volume ratios. For both cases, the length 
of all the nanowires has been chosen as two times that of the cross-section width i.e., an aspect ratio of 2:1.

\begin{table}[h]
\caption{The nanowire size and surface area to volume ratio of different shapes in Scenario-1. In this scenario, the nanowire size 
has been chosen in such a way that all shapes can be inscribed in circular shape as shown in Figure \ref{Fig01}.}
\vspace{0.2cm}
\label{Dimensions}
\centering
\scalebox{0.85}{
\begin{tabular}{| c | c | c |} \hline
\rule{0pt}{3ex}
Nanowire shape & Side length (a), nm & Surface area to volume ratio\\ [3pt]
\hline
\rule{0pt}{3ex}
Triangle & 8.66 & 0.8\\ [3pt]
\rule{0pt}{3ex}
Square & 7.07 & 0.565\\ [3pt]
\rule{0pt}{3ex}
Pentagon & 5.88 & 0.496\\ [3pt]
\rule{0pt}{3ex}
Hexagon & 5 & 0.46\\ [3pt]
\rule{0pt}{3ex}
Circle & 10 & 0.4\\  [3pt]
\hline 
\end{tabular} }
\end{table}

\begin{table}[h]
\caption{The nanowire size and surface area to volume ratio of different shapes in Scenario-2. In this scenario, the nanowire size 
has been chosen in such a way that all the shapes have same surface area to volume ratio.}
\vspace{0.2cm}
\label{Dimensions1}
\centering
\scalebox{0.85}{
\begin{tabular}{| c | c | c |} \hline
\rule{0pt}{3ex}
Nanowire shape & Side length (a), nm  & Surface area to volume ratio\\ [3pt]
\hline
\rule{0pt}{3ex}
Triangle & 17.32 & 0.4 \\ [3pt]
\rule{0pt}{3ex}
Square & 10 & 0.4 \\ [3pt]
\rule{0pt}{3ex}
Pentagon & 7.26 & 0.4 \\ [3pt]
\rule{0pt}{3ex}
Hexagon & 5.77 & 0.4 \\ [3pt]
\rule{0pt}{3ex}
Circle & 10 & 0.4 \\  [3pt]
\hline 
\end{tabular} }
\end{table}
 
Following the creation of the model nanowires with different shapes, the energy minimization was performed by a conjugate 
gradient method to obtain a relaxed structure. To retain the sample at the required temperature, all the atoms have been 
assigned initial velocities according to the Gaussian distribution. Following this, the nanowires were thermally equilibrated 
to a required temperature of 10 K in canonical ensemble (constant NVT) with a Nose-Hoover thermostat. The velocity verlet 
algorithm has been used to integrate the equations of motion with a time step of 2 fs. Following thermal equilibration, the 
tensile deformation was carried out by maintaining a constant strain rate of $1 \times 10^{8}$ s$^{-1}$ along the axial direction. 
Strain ($\varepsilon$) has been calculated as $(l-l_0)/l_0$, where $l$ is instantaneous length and $l_0$ is the initial length 
of the nanowire. The stress has been obtained using the Virial expression \cite{Virial}, which is equivalent to a Cauchy's 
stress in an average sense. Following the MD simulations, the atomic configuration of nanowires during the deformation has 
been visualized using AtomEye package \cite{AtomEye}. Common neighbour analysis (CNA) \cite{CNA1,CNA2} has been used to identify 
the stacking faults.

\section{Results}

The effect of shape on the mechanical properties and deformation behaviour of nanowires has been studied for two different 
scenarios. In the first scenario, the nanowires having different shapes leading to different surface area to volume ratio 
have been chosen (Table \ref{Dimensions}), while in the second scenario, all the nanowires of different shapes have a same 
surface area to volume ratio (Table \ref{Dimensions1}). The results for these two different cases are presented in the 
following; 

\subsection{Nanowires of different shape having different surface area to volume ratio}

\begin{figure}[h]
\centering
\includegraphics[width=9.5cm]{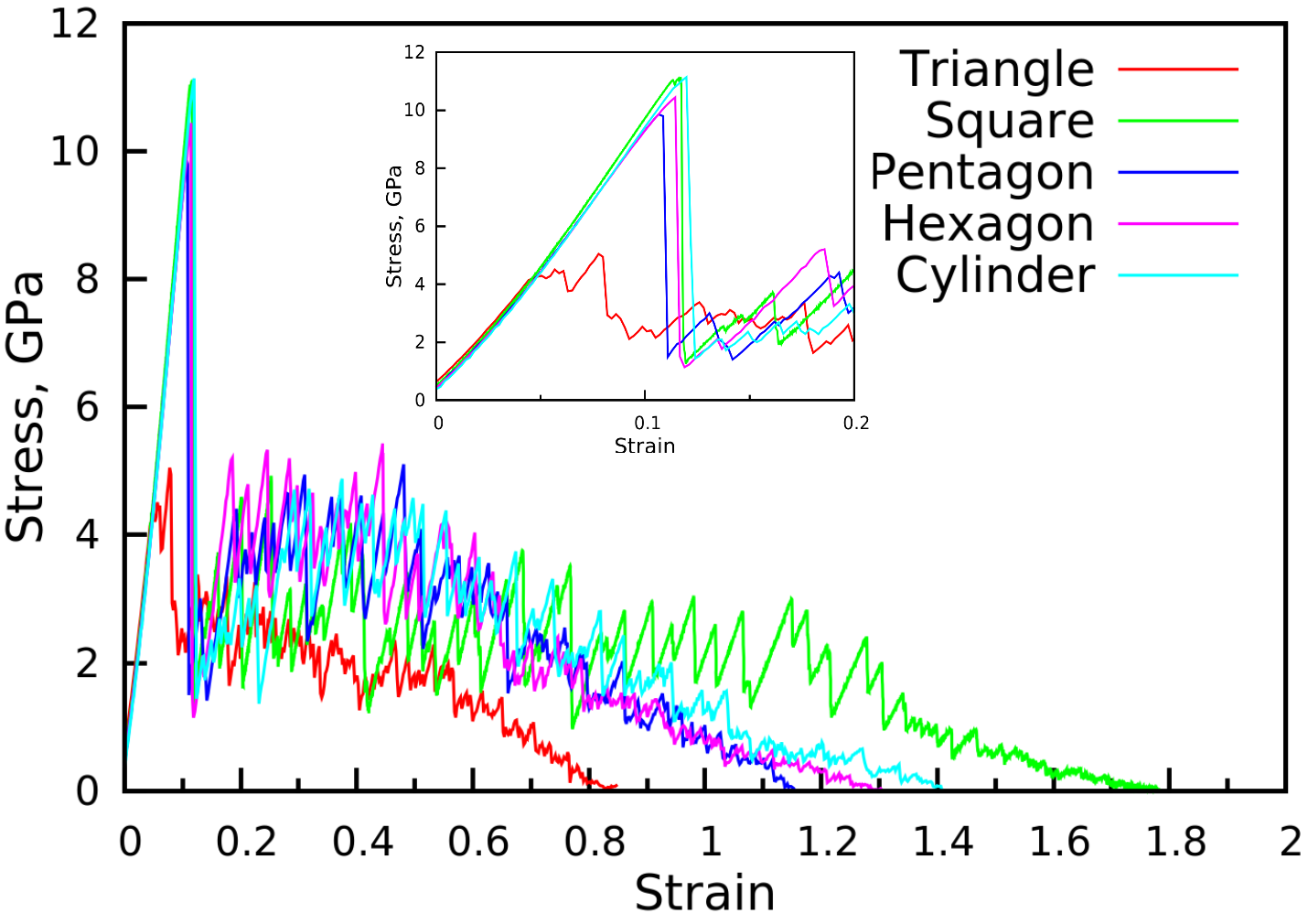}
\caption {Stress-strain behaviour of $<$100$>$ Cu nanowires of different shape in Scenario-1, where each shape has a different 
surface area to volume ratio. The stress-strain behaviour till strain of 20 \% has been shown in inset for better 
clarity.}
\label{Stressstrain_inscribed}
\end{figure}

In this scenario, the size of the nanowires of different shapes has been chosen in such a way that all the shapes can 
be inscribed in nanowire of circular cross-section shape having diameter of 10 nm (Figure \ref{Fig01}). Accordingly, 
all nanowires have different surface area to volume ratio (Table \ref{Dimensions}). The circular nanowire has a lowest 
surface area to volume ratio of 0.4, while the triangular nanowire has the highest and twice the value of circular 
nanowire. Figure \ref{Stressstrain_inscribed} shows the stress-strain behaviour of nanowires of different shape having 
different surface area to volume ratio. For better clarity, inset figure shows the stress-strain curve till a strain 
value of 20\%. Initially, all the nanowires show linear elastic behaviour up to a peak value followed by an abrupt drop 
in flow stress (except triangular nanowire). In triangular nanowire, the abrupt drop in flow stress is absent and flow 
stress increases marginally after yielding to a peak value of 5.16 GPa. Following the yielding in all the nanowires, 
the flow stress decreases continuously with increase in strain till failure. During this decrease, the flow stress 
exhibits a rapid fluctuations (Figure \ref{Stressstrain_inscribed}). Further, the flow stress of the triangular 
nanowires is always lower than the flow stress of the other nanowires. Finally, the square shaped nanowire exhibits 
the highest failure strain of around 180\%, while the failure strain is lowest for triangular shaped nanowire.

\begin{figure}
\centering
\includegraphics[width=9.5cm]{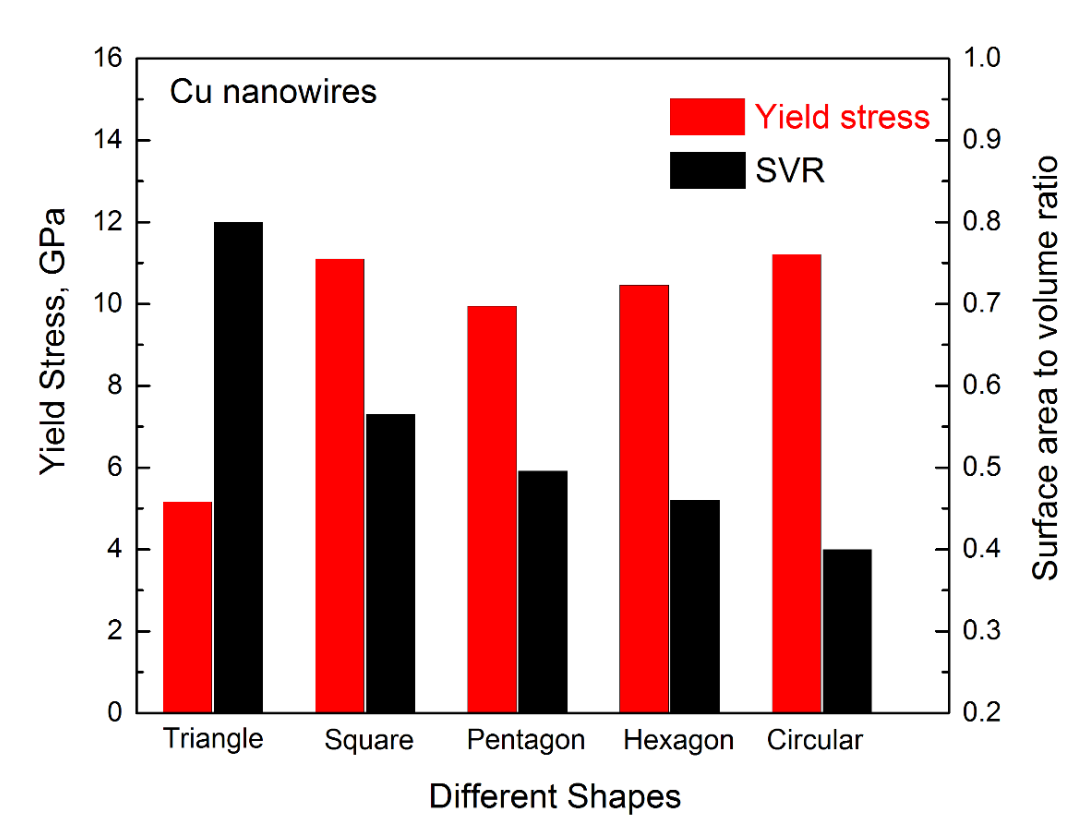}
\caption {Variation of yield strength along with surface area to volume ratio for different nanowire shapes in Scenario-1.}
\label{Yieldstress-NSVR}
\end{figure}

Figure \ref{Stressstrain_inscribed} also shows that the slope of linear elastic regime is almost same for all the 
nanowires, which indicates that the Young's modulus is insensitive to the nanowire shape. The value of Young's modulus 
is obtained as 92 GPa, which is in good agreement with previous reported value \cite{Liang2005}. Further, in the 
atomistic simulation study of Cao and Ma \cite{Cao2008}, it was observed that both circular and square shaped nanowires 
possess the same value of Young's modulus, which affirm the present observation that the Young's modulus is insensitive 
to nanowire shape. Here, it is interesting to see that despite the different surface area to volume ratio, Young's 
modulus remains same for different shapes. However, it must be noted that Young's modulus can be influenced by surface 
area to volume ratio when it is varied by nanowire size \cite{Rohith-Philmag, Rohith-CMS17, Liang2005}. The variations 
in yield strength for nanowires of different shapes has been presented in Figure \ref{Yieldstress-NSVR}. It can be seen 
that the triangular nanowires exhibits a minimum yield strength value of 5.16 GPa. The magnitude of yield strength 
increases with increasing the number of the side surfaces and attains a maximum for circular shaped nanowire (Figure 
\ref{Yieldstress-NSVR}). The variations in yield stress values for different shapes can be attributed to the variations 
in surface area to volume ratio. The surface area to volume ratio for different shapes has been calculated (Table 
\ref{Dimensions}) and plotted by superimposing in Figure \ref{Yieldstress-NSVR}. It can be seen that the surface area 
to volume ratio decreases in the order of triangular, pentagonal, hexagonal, and circular nanowires. These 
variations clearly indicates that yield stress is inversely proportional to surface area to volume ratio, i.e., the 
yield strength of nanowires is high for nanowires with low surface area to volume ratios (Figure \ref{Yieldstress-NSVR}). 

\subsection{Nanowires of different shapes having same surface area to volume ratio}

\begin{figure}[h]
\centering
\includegraphics[width=9.5cm]{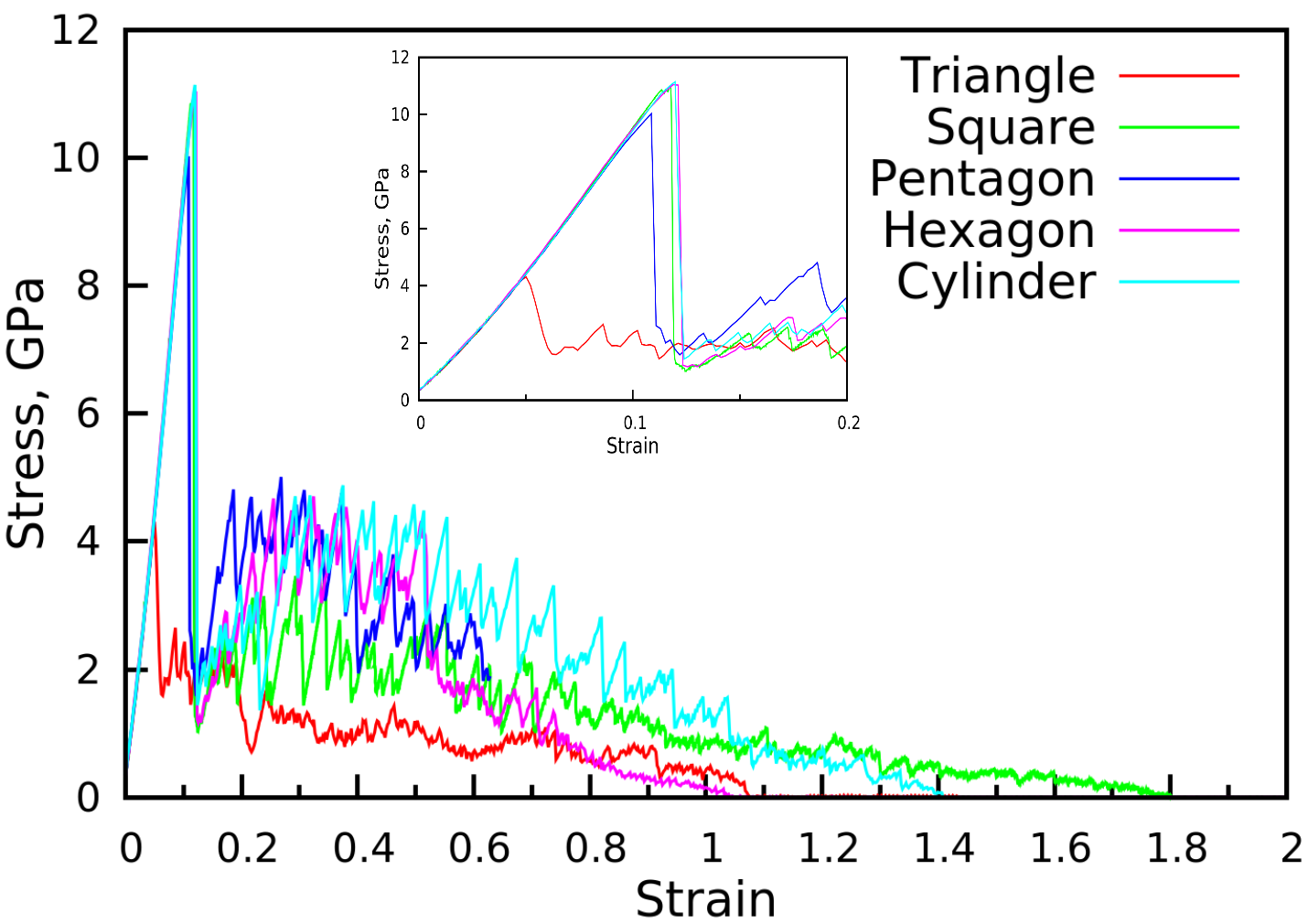}
\caption {Stress-strain behaviour of $<$100$>$ Cu nanowires of different shape in Scenario-2, where all the shapes have 
same surface area to volume ratio. The stress-strain behaviour till strain of 20 \% has been shown in inset for better 
clarity. }
\label{Stressstrain_CSVR}
\end{figure}

In previous section, the effect of shape has been studied by considering different nanowire shapes having different surface 
area to volume ratio. In order to study the sole effect of shape, here the surface area to volume ratio has been kept same 
for all shapes. Accordingly, the size of the nanowires has been varied as shown in Table. \ref{Dimensions1}. Figure 
\ref{Stressstrain_CSVR} shows the stress-strain behaviour of nanowires of different shapes with same surface area to volume 
ratio under tensile loading. For clarity, the magnified version of stress-strain behaviour till a strain value of 20 $\%$ 
has been shown in inset figure. Similar to scenario-1, all the nanowires exhibit linear elastic deformation up to a peak 
value of stress. After elastic deformation, all the nanowires including the triangular one show a drastic yield drop followed 
by decrease in flow stress with increasing strain until failure. Further, as in the previous case, the flow stress of the 
triangular nanowires is considerably lower than the flow stress of the other nanowires (Figure \ref{Stressstrain_CSVR}). 
Finally, the square shaped nanowire exhibits the highest failure strain, while it is lowest for hexagon and triangular 
shaped nanowires (Figure \ref{Stressstrain_CSVR}).

\begin{figure}[h]
\centering
\includegraphics[width=9.5cm]{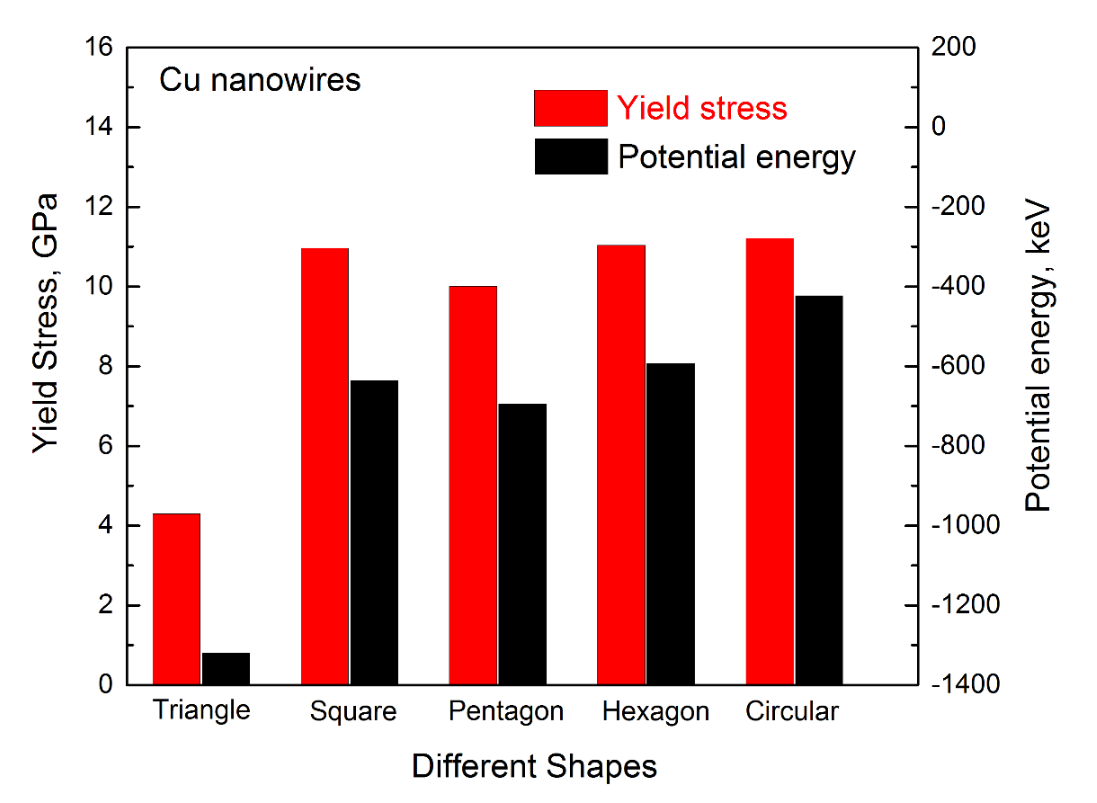}
\caption {Variation of yield strength and potential energy for different nanowire shapes in Scenario-2.}
\label{Yieldstress-pot_CSVR}
\end{figure}

Similar to previous scenario, the Young's modulus has been calculated as the slope of the linear elastic regime and found to 
be around 92 GPa for all nanowires. This indicate that the Young's modulus is insensitive to the nanowire shape, irrespective 
of whether the surface area to volume ratio is varied or kept constant. The variations in yield stress for different shaped 
nanowires with same surface area to volume ratio is shown in Figure \ref{Yieldstress-pot_CSVR}. Here, it is interesting to 
observe that despite the same surface area to volume ratio, the yield strength still varies for different shapes. Like the 
previous scenario, the triangular nanowire exhibits the lowest yield stress and it increases with increasing number of 
side surfaces before attaining the maximum for circular nanowire. However, unlike the previous case, this variation in 
yield stress cannot be explained by the surface area to volume ratio as it is constant. To understand this variations in 
the present scenario, the potential energy for different shapes has been calculated and shown in Figure \ref{Yieldstress-pot_CSVR}. 
Interestingly, both the potential energy and yield stress varies in a similar fashion. Like yield stress, the potential 
energy also increases with increase in number of side surfaces exhibiting the highest value for circular nanowire and the 
lowest for triangular nanowire. These variations indicates that there exists a one to one correlation between potential 
energy and yield stress. Hence, the variations in yield strength can be attributed to the 
variations in potential energy of the nanowires (Figure \ref{Yieldstress-pot_CSVR}). Here, it must be noted that in the 
previous case (nanowires with different shape and different surface area to volume ratio), there were no significant 
variations in potential energy and no correlation has been found between yield stress and potential energy.

\section{Deformation behaviour}

\begin{figure}[h]
\centering
\includegraphics[width=10cm]{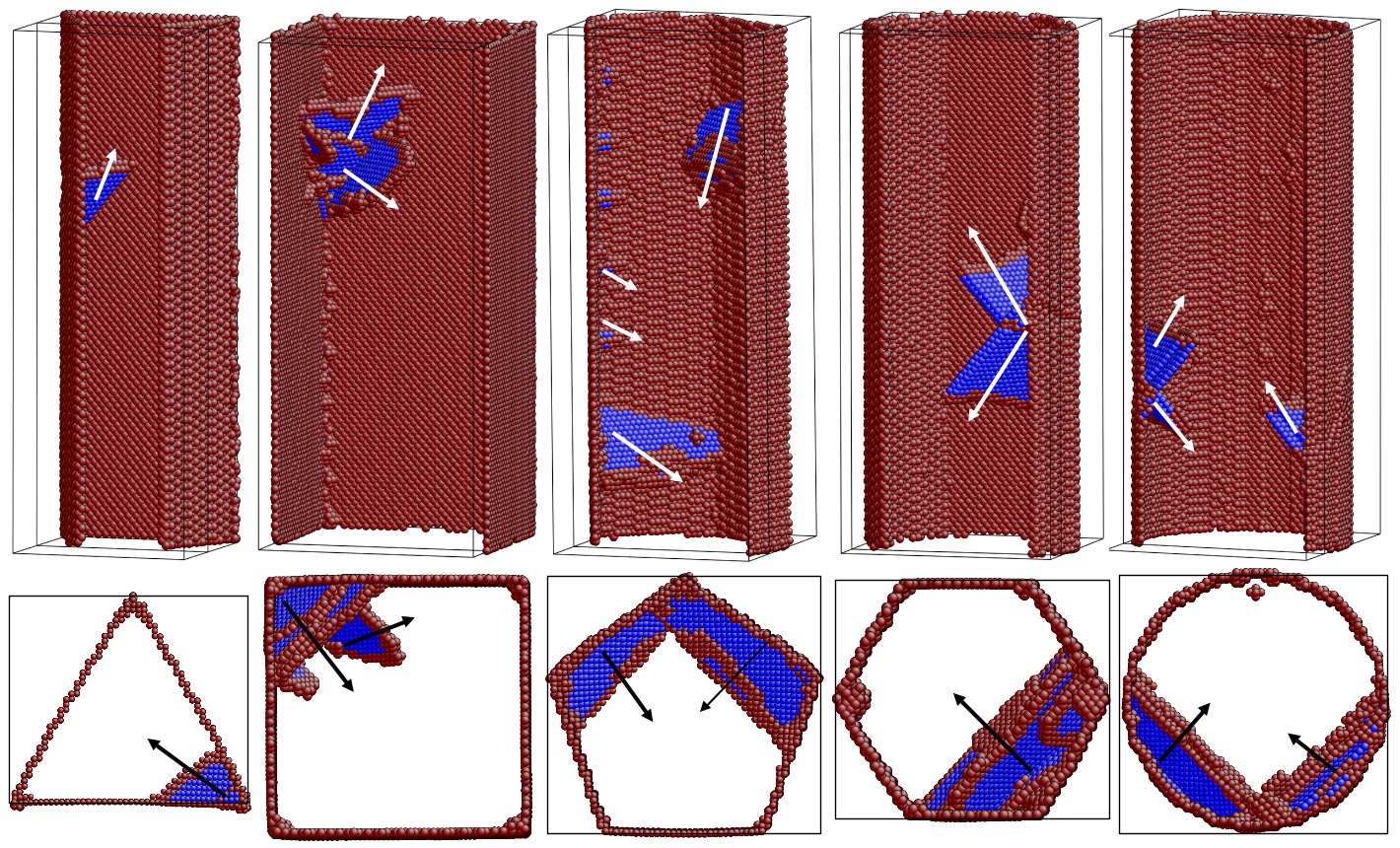}
\caption {The dislocation nucleation during the yielding/beginning of plastic deformation in nanowires of different shapes. 
This nucleation mechanism remains the same for both scenarios. The white arrows indicates the direction of dislocation 
motion. The nucleation occurs from the intersection of two different side surfaces.}
\label{Nucleation}
\end{figure}

In order to understand the effect of shape on the dislocation nucleation and deformation mechanisms, the atomic configurations 
have been analysed for two different scenarios using the CNA parameter and visualized using AtomEye. The results indicate that 
the defect nucleation and deformation mechanisms were same for both the scenarios. Following yielding, the sudden drop in flow 
stress corresponds to the defect nucleation in the nanowires and this can also be considered as the initiation of plastic 
deformation. The initial defect nucleation corresponding to the yielding in the nanowires of different shapes is shown in 
Figure \ref{Nucleation}. It can be seen that in all the nanowires including the circular one, the yielding occurs through the 
nucleation of 1/6$<$112$>$ Shockley partial dislocations from the corners (or intersection of two different side surfaces) 
of the nanowires. In circular nanowire, the defect nucleation has been observed from the intersection of \{212\} and \{110\} 
type side surfaces (Figure \ref{Fig01}). It is well known that the atoms at the corners (or intersection of two different side 
surfaces) of nanowires have low coordination resulting in high stress concentration. As a result, the corners act as a favourable 
sites for defect nucleation (Figure \ref{Nucleation}). 

\begin{figure}[h]
\centering
\includegraphics[width=10cm]{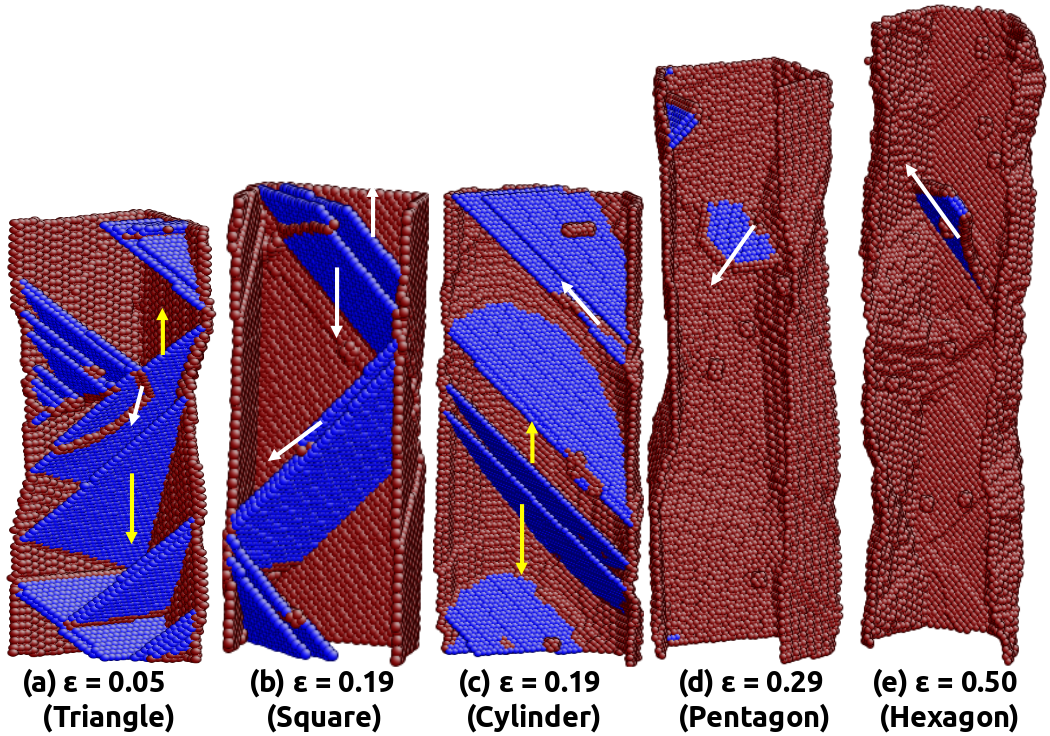}
\caption {The deformation mechanisms observed in nanowires of different shapes, each shown at different strain value. The yellow 
arrows indicate the direction of twin boundaries movement. The white arrows indicate the direction of glide of Shockley partial 
dislocations on \{111\} planes.}
\label{Deformation-allshapes}
\end{figure}

\begin{figure}[h]
\centering
\includegraphics[width=7.5cm]{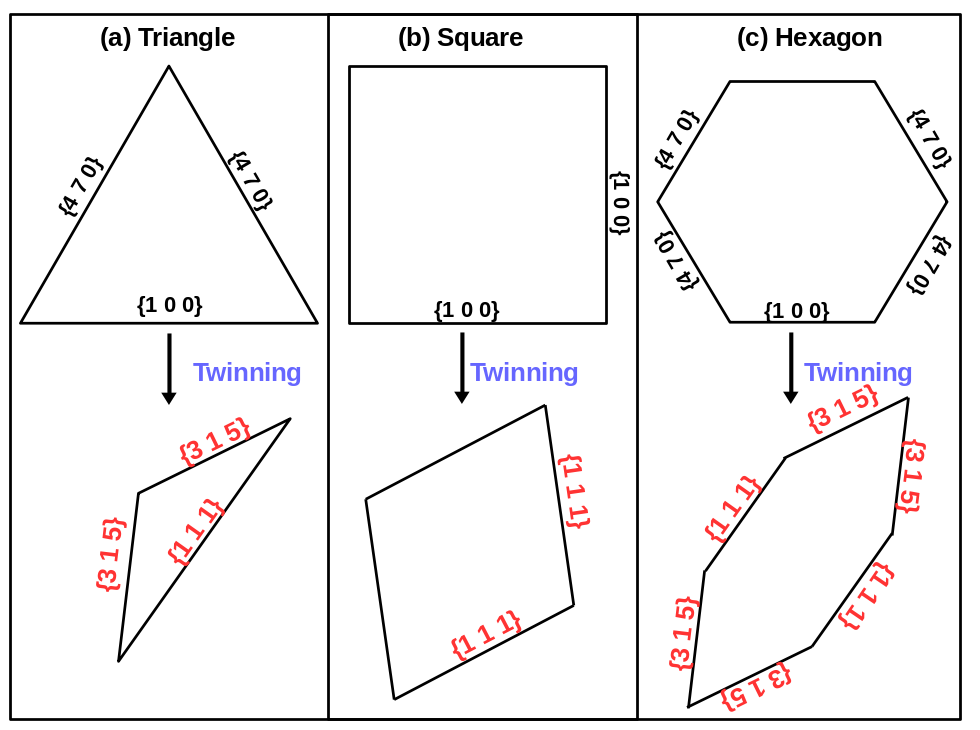}
\caption {The schematic showing the orientation of initial (top row) and twinned side surfaces for (a) triangular, (b) square and, 
(c)hexagonal shaped nanowires. In pentagonal and circular nanowires, eventhough the deformation twinning has been observed, but we 
are not able to identify the new surface orientations due to high roughness.}
\label{twinsurfaces}
\end{figure}

Following the nucleation of Shockley partials from the corners, the deformation behaviour in nanowires of 
different shapes is dominated by the slip through partial dislocations along with deformation twinning (Figure 
\ref{Deformation-allshapes}). However, occasionally the extended dislocations were also observed, but only at high strains 
($>50\%$). As shown in Figure \ref{Deformation-allshapes}, the twinning has been observed on multiple \{111\} planes. It has 
been well known that the deformation twinning alters the surface orientation of the nanowires at the twinned region 
\cite{Park2006,Weinberger2012Review,Rohith-ComCondMater,Dia02004}. In the present study, the deformation twinning changed the 
initial \{100\} side surface of square shaped nanowires to \{111\} type surface (Figure \ref{twinsurfaces}b). Similarly, in 
triangular and hexagonal shaped nanowires, the twinning changes the initial \{470\} surface to \{315\} type surface and \{100\} 
surface to \{111\} surface (Figure \ref{twinsurfaces}a \& c). In pentagonal and circular nanowires, eventhough the deformation 
twinning has been observed, but we are not able to identify the new surface orientations due to high roughness.

\begin{figure}[h]
\centering
\includegraphics[width=7.5cm]{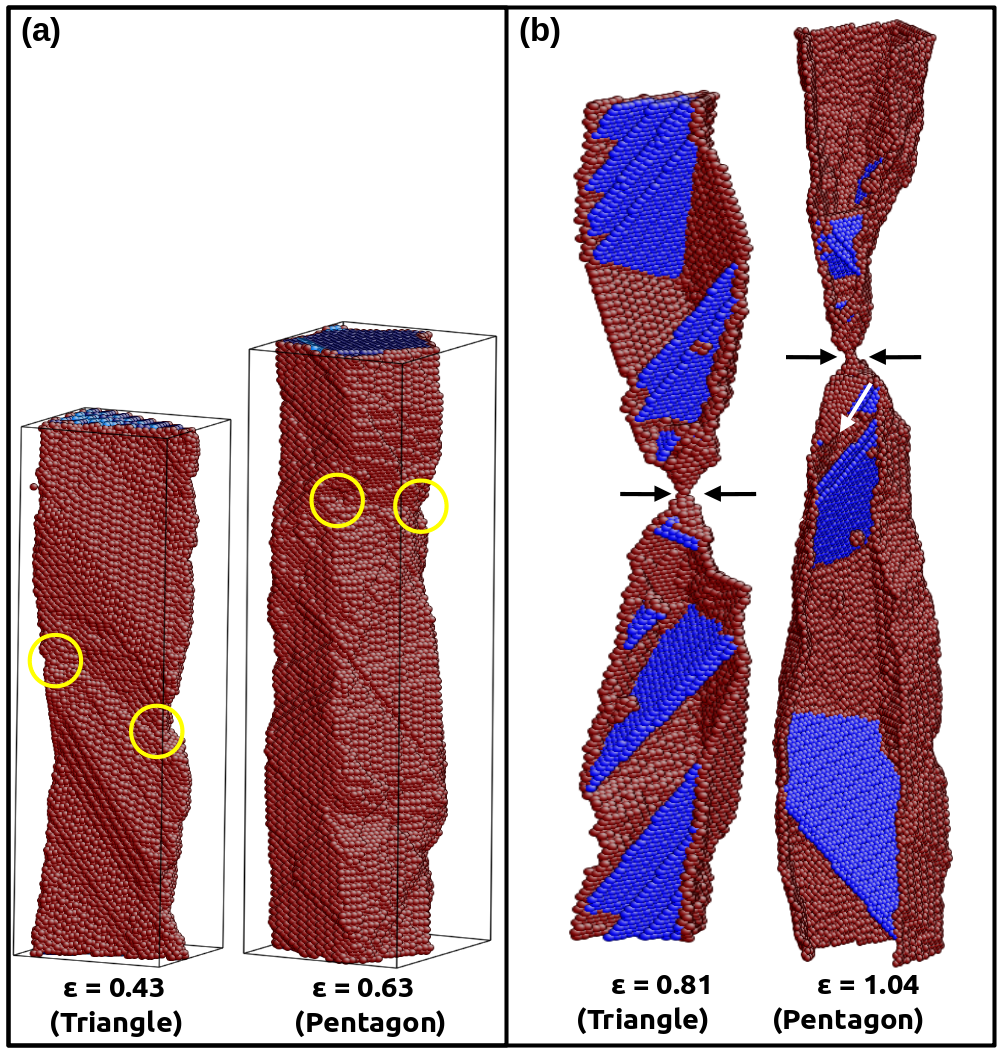}
\caption {The final failure behaviour of nanowires of different shapes. The encircled regions indicate the initian of necking 
and black arrows shows failed region.}
\label{Failure-allshapes}
\end{figure}

Figure \ref{Failure-allshapes} shows the necking and final failure behaviour in nanowires with different cross-section shapes. 
As shown in Figure \ref{Deformation-allshapes}, the plastic deformation spreads across the nanowire length. Following uniform 
deformation, the discrete dislocation interactions among different partials leads to the formation of a localized necking 
(Figure \ref{Failure-allshapes}). In all nanowires, like dislocation nucleation, the necking starts from the corners or the 
intersection of different side surfaces of the nanowires (Figure \ref{Failure-allshapes}). The formation of necking localizes 
the plastic deformation and also minimize the effect of corners for any further dislocation nucleation. The localized plastic 
deformation results in final failure of the nanowires in a ductile manner (Figure \ref{Failure-allshapes}).

In single crystal nanowires, the deformation by partial or full dislocation slip can be understood based on the parameter 
$\alpha_M$, defined as the ratio of Schmid factors for leading and trailing partial dislocations \cite{Rohith-ComCondMater,TB-sliding}. 
For a given orientation of a nanowire, if $\alpha_M$ $>1$, the deformation by twinning/partial dislocation slip is favoured. 
On the other hand, if $\alpha_M$ $<1$, the deformation by full/extended dislocations is favoured. Similarly, if $\alpha_M$ =1, 
the deformation can proceed either through twinning or extended dislocations or by the combination of these two. It has been 
shown that for $<100>$ orientation, the value of $\alpha_M$ is 0.5 under tensile loading \cite{Rohith-ComCondMater}. Accordingly, 
the deformation by full/extended dislocation should be observed. However, in the present study the deformation through slip 
of Shockley partials has been observed (Figure \ref{Deformation-allshapes}). This discrepancy between the predicted and observed 
deformation mechanism can be attributed to the effects associated with side surfaces \cite{Park2006, Weinberger2012Review}, whose 
contribution is absent in the analysis based on $\alpha_M$. Similar to the present study, many previous studies 
\cite{Park2006,Weinberger2012Review} have shown that the nanowires oriented $<100>$ tensile axis deform by partial dislocation 
slip under tensile loading.

\section{Conclusions}

Molecular dynamics simulations have been performed to understand the influence of nanowire shape on the mechanical properties and the 
deformation mechanisms of Cu nanowires. Five different shapes (triangular, square, pentagon, hexagonal and circular) have been 
considered. Based on nanowires size, two different scenarios have been chosen. In the first scenario, the size of the nanowires 
has been chosen in such a way that all the shapes can be inscribed in nanowire of circular cross-section with a diameter 10 nm.
In this case, all nanowires possess different surface area to volume ratios. In the second scenario, the size of the nanowires 
has been chosen in such a way that the surface area to volume ratio remains same for all shapes. The MD simulations results 
indicate that the Young's modulus in both scenarios is insensitive to the nanowire shape. This insensitivity of Young's modulus 
is interesting particularly in first scenario, where different shapes have different surface area to volume ratio. In both the 
scenarios, the yield strength increases with increasing number of side surfaces. The traingular nanowire shows the lowest strength 
and ductility (failure strain), while the circular nanowire display highest strength combined with good ductility. Here, it is 
interesting to observe that despite the same surface area to volume ratio in the second scenario, the yield strength still varies 
for different shapes. The variations in yield strength have been explained based on the variations in surface area to volume 
ratio (First scenario) and potential energy (Second scenario) of the nanowires. In all the nanowires, irrespective of their 
shape and surface area to volume ratio, the deformation is dominated by the slip through Shockley partial dislocations and 
deformation twinning. The deformation twinning changed the orientation of side surfaces at the twinned region, whose orientation 
is different for different shapes. Finally nanowires of all shapes fail in a ductile manner.


\section*{References}

\begin{thebibliography}{99}

\bibitem{Lieber2003} C.M. Lieber, Nanoscale science and technology: Building a big future from small things, MRS Bull. 28 (2003) 486.

\bibitem{Review-1} Y. Chen, X. An, and X. Liao, Mechanical behaviors of nanowires, Appl. Phys. Rev. 4 (2017) 031104. 

\bibitem{Review-2} S. Wang, Z. Shan, and H. Huang, The mechanical properties of nanowires, Adv. Sci. 4 (2017) 1600332.

\bibitem{Uchic} M.D. Uchic, D.M. Dimiduk, J. N. Florando, W.D. Nix, Sample dimensions influence strength and crystal plasticity, 
Science 305 (2004) 986.

\bibitem{Peng2012} C. Peng, Y. Zhan, J. Lou, Size-dependent fracture mode transition in copper nanowires, Small 8 (2012) 1889.

\bibitem{Yaghoobi2016} M. Yaghoobi, G.Z. Voyiadjis, Size effects in fcc crystals during the high rate compression test, Acta Mater. 
121 (2016) 190.

\bibitem{Rohith-Philmag} G. Sainath, P. Rohith, B.K. Choudhary, Size dependent deformation behaviour and dislocation mechanisms in 
$<100>$ Cu nanowires, Philos. Mag. 97 (2017) 2632.

\bibitem{Rohith-CMS17} P. Rohith, G. Sainath, B.K. Choudhary, Molecular dynamics simulation studies on the influence of aspect ratio 
on tensile deformation and failure behaviour of $<100>$ copper nanowires, Comp. Mater. Sci. 138 (2017) 34.

\bibitem{Zhu2012} Y. Zhu, Q. Qin, F. Xu, F. Fan, Y. Ding, T. Zhang, B. J. Wiley, and Z.L. Wang, Size effects on elasticity, yielding, 
and fracture of silver nanowires: In situ experiments, Phys. Rev. B 85 (2012) 045443. 

\bibitem{Park2006} H.S. Park, K. Gall, J.A. Zimmerman, Deformation of FCC nanowires by twinning and slip J. Mech. Phys. Solids 54 
(2006) 1862.

\bibitem{Weinberger2012Review} C.R. Weinberger and W. Cai, Plasticity of metal nanowires, J. Mater. Chem. 22 (2012) 3277.

\bibitem{Rohith-ComCondMater} P. Rohith, G. Sainath, B.K. Choudhary, Effect of orientation and mode of loading on deformation 
behaviour of Cu nanowires Comp. Condens. Mat. 17 (2008) e00330. 
 
\bibitem{Xie2015} H. Xie, F. Yin, T. Yu, G. Lu, Y. Zhang, A new strain-rate-induced deformation mechanism of Cu nanowire: Transition 
from dislocation nucleation to phase transformation, Acta Mater. 85 (2015) 191.

\bibitem{Cao2008} A. Cao, E. Ma, Sample shape and temperature strongly influence the yield strength of metallic nanopillars. Acta 
Mater. 56 (2008) 4816.

\bibitem{Sai-JAP} G. Sainath, B. K. Choudhary, Atomistic simulations on ductile-brittle transition in $<$111$>$ BCC Fe nanowires, 
J. Appl. Phys. 122 (2017) 095101. 

\bibitem{Leach2007} A.M. Leach, M. McDowell, K. Gall, Deformation of top-down and bottom-up silver nanowires, Adv. Funct. Mater. 
17 (2007) 43.

\bibitem{Zhang2009} Y. Zhang, H. Huang, Do Twin Boundaries Always Strengthen Metal Nanowires?, Nano Res. Lett. 4 (2009) 34. 

\bibitem{Arora2017} N. Arora, D. P. Joshi, P. Uma, Size and shape dependent Debye temperature of nanomaterials. Mater. Today Proc. 
4 (2017) 10450.

\bibitem{Kateb2018} M. Kateb, M. Azadeh, P. Marashi, S.Ingvarsson, Size and shape-dependent melting mechanism of Pd nanoparticles, 
J. Nanopart. Res. 20 (2018) 251.

\bibitem{Kadam2019} K.D. Kadam, S.L. Patil, H.S. Patil, P.P. Waifalkar, K.V. More, R.K. Kamat, T.D. Dongale, Shape dependent 
optical properties of GaAs quantum dot: A simulation study, J. Nano Electr. Phys. 11 (2019) 01013.


\bibitem{Plimpton-1995} S. Plimpton, Fast parallel algorithms for short-range molecular dynamics, J. Comp. Phy. 117 (1995) 1.

\bibitem{Mishin-2001} Y.Mishin, M.J. Mehl, D.A. Papaconstantopoulos, A.F. Voter, J.D. Kress,  Structural stability and lattice 
defects in copper: Ab initio, tight-binding, and embedded-atom calculations, Phys. Rev. B, 63 (2001) 1.

\bibitem{Liang-PRB} W. Liang, M. Zhou, Atomistic simulations reveal shape memory of fcc metal nanowires, Phys. Rev. B 73 (2006) 
115409.

\bibitem{Virial} J.A. Zimmerman, E.B. Webb, J.J. Hoyt, R.E. Jones, P.A. Klein, D.J. Bammann,  Calculation of stress in atomistic 
simulations, Modell. Simul. Mater. Sci. Eng. 12 (2003) S319.

\bibitem{AtomEye} J. Li, AtomEye: an efficient atomistic configuration viewer, Modell. Simul. Mater. Sci. Eng. 11 (2003) 173.

\bibitem{CNA1} D. Faken, H. Jonsson, Systematic analysis of local atomic structure combined with 3D computer graphics, Comp. Mater. 
Sci. 2 (1994) 279.

\bibitem{CNA2} H. Tsuzuki, P.S. Branicio, J.P. Rino, Structural characterization of deformed crystals by analysis of common atomic 
neighborhood, Comp. Phy. Comm. 177 (2007) 518.

\bibitem{Liang2005} H. Liang, M. Upmanyu, H. Huang, Size-dependent elasticity of nanowires: Nonlinear effects, Phys. Rev. B 71 
(2005) 241403 (R).

\bibitem{Dia02004} J. Diao, K. Gall, M.L. Dunn, Surface stress driven reorientation of gold nanowires, Phys. Rev. B 70 (2004) 
075413. 

\bibitem{TB-sliding} Z.J. Wang, Q.J. Li, Y. Li, L.C. Huang, L. Lu, M. Dao, J. Li, S. Suresh, Z.W. Shan, Sliding of coherent twin 
boundaries, Nat. Commun. 8 (2017) 1108.

\end{thebibliography}

\end{document}